# Enhancing the resolution of hyperlens by the compensation of losses without gain media

Xu Zhang[1], Wyatt Adams[1], Mehdi Sadatgol[1], and Durdu Ö. Güney[1], *

[1]Department of Electrical and Computer Engineering, Michigan Technological University, 1400 Townsend Dr., Houghton, MI 49931-1295, USA
*Corresponding author: dguney@mtu.edu

We present a method to improve the resolution of available hyperlenses in the literature. In this method, we combine the operation of hyperlens with the recently proposed plasmon injection scheme for loss compensation in metamaterials. Image of an object, which is otherwise not resolvable by the hyperlens alone, was reconstructed up to the minimum feature size of one seventh of the free-space wavelength.

## I. INTRODUCTION

Due to the direct control and manipulation of electromagnetic properties, metamaterials provide previously unthought-of approaches for high resolution imaging [1-9], high efficiency photovoltaics [10], and novel optical materials [11-14], among others. In 1873, Abbe discovered that when an object feature is smaller than half of the wavelength of the light, it cannot be resolved by conventional optics because of diffraction [15]. Metamaterials provide the possibility to overcome the diffraction limit. Perfect lenses [1,2], superlenses [3,4], and hyperlenses [5-9] have been theoretically proposed and fabricated. Among them, hyperlenses have emerged as one of the most interesting and promising lenses due to their ability to propagate a sub-diffraction-limited image into the far-field. Hyperlenses are made of hyperbolic metamaterials [16,17] which convert ordinary evanescent waves (corresponding to subwavelength features) into propagating waves that can be imaged by a conventional lens in the far-field.

Unfortunately, the high absorptive losses [18] in the constitutive components limit the hyperlens resolution. The smallest feature which can be resolved so far experimentally is around $\lambda_0/3$ [9,19], where $\lambda_0$ is the free-space wavelength. Recently, a loss compensation scheme called plasmon injection ($\Pi$) scheme was proposed. The $\Pi$ scheme relies on the coherent superposition of externally injected surface plasmon polaritons with the local eigenmodes of a metamaterial to provide full loss compensation [20-22]. This technique does not need traditional optical-gain providing medium [23,24], hence eliminates its associated complexities, and is more importantly equivalent to applying a simple spatial filter for imaging [22].

In this Letter, we use the design of one experimentally realized cylindrical hyperlens as an example to demonstrate the applicability of this technique to hyperlensing for higher resolution imaging.

The hyperlens we studied here is from [19]. To begin the procedure, the simulation result from [19] is replicated using the commercial finite element solver COMSOL Multiphysics. Fig. 1 shows the simulated magnetic field distribution. The hyperlens consists of 8 pairs of concentric Ag/Al$_2$O$_3$ layers, with the surrounding material being quartz. The thickness of the Ag and Al$_2$O$_3$ layers is 35nm. At $\lambda_0 = 365nm$ working wavelength, the permittivities of Ag, Al$_2$O$_3$ and quartz are $\varepsilon_m = -2.4012 + 0.2488i$, $\varepsilon_d = 3.217$, and $\varepsilon_{qtz} = 2.174$, respectively. The hyperlens is illuminated by a transverse-magnetic polarized plane wave (i.e., magnetic field is along the axis of the cylindrical hyperlens) using a port backed by perfectly matched layers (PML) to absorb outgoing waves. The result is in agreement with Fig. 2 (b) from [19].

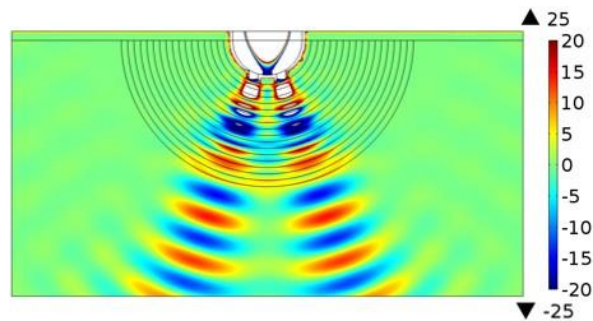

FIG. 1. Replicated hyperlens magnetic field (A/m) distribution simulation result corresponding to Fig. 2 (b) from [19]. Two 50nm wide openings in a 50nm thick Cr layer are considered as the object. The center to center separation of the openings is 150nm. The working wavelength is 365nm.

After verifying the simulation result, the imaging process can begin as described here. First, the raw image is obtained by the hyperlens. Due to absorptive loss in the hyperlens, the high spatial frequency components of the object are attenuated on the image plane. Then, a filter is applied to compensate this attenuation. After this post processing, a high resolution image will be obtained. The compensation filter applied here is the inverse of the hyperlens transfer function, which is calculated by simulation. Interestingly, this corresponds to recently proposed $\Pi$ scheme loss compensation technique for imaging [22]. In the $\Pi$ scheme, the total incident field in the object plane is a coherent superposition of the main object to be imaged and some auxiliary object. The auxiliary object coherently excites the underlying modes of the system, resulting in a compensation of the attenuation in the main object. This scheme is equivalent to applying a filter in the Fourier domain to amplify the high spatial frequency components. Although inverse filtering is well-known for propagating modes in the field of image processing, application to evanescent modes and intimate relation with loss compensation distinguish the work presented here from traditional inverse filtering. Similar inverse filtering approach to countering losses in Ag superlens and negative index flat lenses has been recently considered in [22,25,26].

## II. RESULTS AND DISCUSSIONS

Before showing the procedure for calculation of the hyperlens transfer function, the object plane and image plane should be defined. Consider that the object plane is defined at the inner face of the hyperlens. Due to the conservation of angular momentum [5], the tangential component $k_\theta$ and, according to the dispersion relation given by $k_r^2/\varepsilon_\theta - k_\theta^2/|\varepsilon_r| = (\omega/c)^2$ in cylindrical coordinates, the radial component $k_r$ of the wave vector decrease as a wave propagates through the hyperlens. For different spatial frequency components, the phase is restored at different positions. Unlike the case of a negative index flat lens [25], the hyperlens needs both amplitude and phase compensation. Fig. 2 shows the geometry for the transfer function calculation using COMSOL. To avoid spherical aberration of the image, a curved image plane is designed. There is no unique image plane, so the location of the image plane d can be arbitrarily chosen.

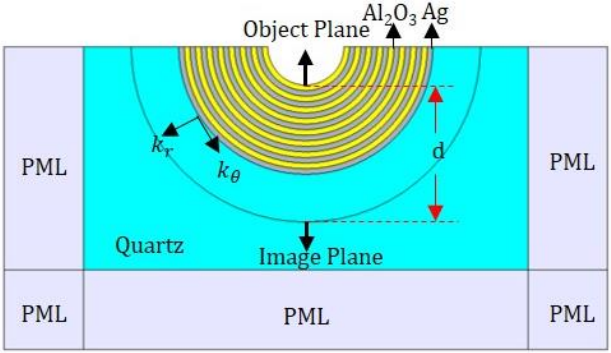

FIG. 2. Geometry for the COMSOL transfer function calculation. The resulting magnetic field image is recorded at a radial distance d from the object plane. Perfectly matched layers are added at the boundary to absorb the outgoing waves without reflecting them back to the interior.

The transfer function is defined as the division of the magnetic field at the image plane and object plane. We selected $d = 860nm$ to show the process of calculating the transfer function, though we note that the transfer function varies with d. The object plane radius (i.e., inner radius of the hyperlens) is $r_{op} = 240nm$ and the image plane radius is $r_{ip} = 1100nm$. Cylindrical waves with different spatial frequency $k_\theta$ were defined at the object plane. The amplitude and phase information of the resulting magnetic field was measured at the image plane. It should be pointed out that for $k_\theta$ at the object plane, the corresponding spatial frequency is changed to $k_\theta'$ at the image plane. The relationship between $k_\theta$ and $k_\theta'$ is given by $k_\theta r_{op} = k_\theta' r_{ip} = m$, where $m$ is the angular momentum mode number of the cylindrical wave [5]. Therefore, the transfer function is a function of m. Fig. 3 shows the transfer function and corresponding compensation filter for amplitude and phase.

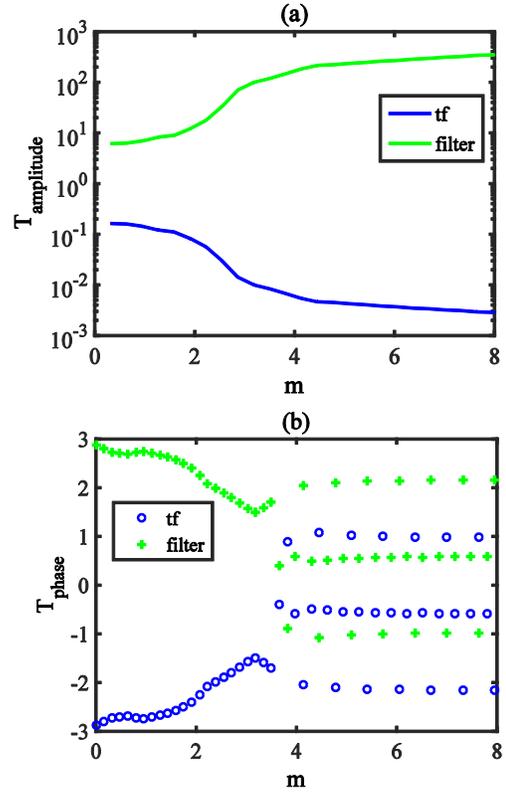

FIG. 3. Calculated (a) amplitude and (b) phase of the hyperlens transfer function (blue) plotted with the corresponding compensation filter (green). Phase data points are not interpolated due to rapid change with $m$.

To demonstrate the imaging procedure, a magnetic field with four Gaussian features is defined at the object plane as an example. Note that this procedure can be performed with any arbitrary features at the object plane. The field is defined so that the smallest separation between two peaks is 50nm. To begin, the raw image is obtained through the hyperlens at the image plane. Then in the spatial frequency domain, the Fourier transform of the raw image is multiplied by the compensation filter. The amplitude and phase compensation results are shown in Figs. 4 (a) and (b), respectively. After transformation back to the spatial domain, the compensated image is obtained in Fig. 5. The image is magnified because of the hyperlens dispersion and geometry. Therefore, in order to clearly show the result, $\theta$ is used as the horizontal axis in Fig. 5. It is clearly seen that the sub-diffraction-limited features of the object can be reconstructed with a resolution of $\lambda_0/7$ after applying the loss compensation filter, while the raw image obtained by the hyperlens alone cannot be resolved.

We should note that at high spatial frequencies, the filter will also amplify the simulation noise, which will influence the compensated image. Therefore, we truncate the filter at $m = 6$, where the noise floor is reached, to avoid a strong influence of noise amplification. The resultant truncated object, which maintains the major sub-diffraction-limited features of the original object, is also shown in Fig. 5. The loss compensation filter almost perfectly reconstructs the truncated object.

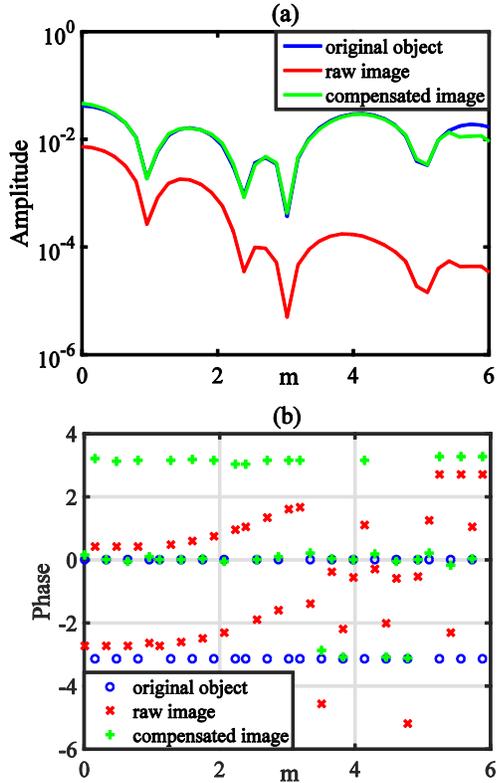

FIG. 4. (a) Amplitude and (b) phase Fourier spectra for the original object, raw image, and compensated image, respectively. Phase data points are not interpolated due to rapid change with $m$.

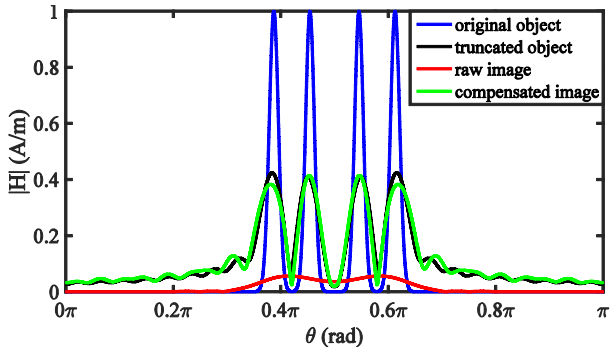

FIG. 5. Magnetic field intensity for the original object, truncated object, raw image, and compensated image. The compensated image is clearly resolved beyond the diffraction limit.

To illustrate the equivalence of the above spatial filtering process with the Π scheme [22], we simulate directly the coherent superposition of an auxiliary object and the original object as an input to the hyperlens and find the resultant image instead of simulating the original object alone and then performing the spatial filtering (see Fig. 6). The superposed total input is calculated by dividing the compensated image in Fig. 5 by the transfer function in Fig. 3. The auxiliary object (not shown) is the difference between the total input and the original object due to the linearity of the system. It is clearly seen in Fig. 6 that the image obtained through the direct simulation of the total input strongly agrees with the image obtained through the spatial filtering. This suggests that the Π scheme in which a part of the total input (i.e., auxiliary object) physically compensates the losses in the hyperlens to leave the original object intact is equivalent to applying mathematically a compensating spatial filter to the raw image.

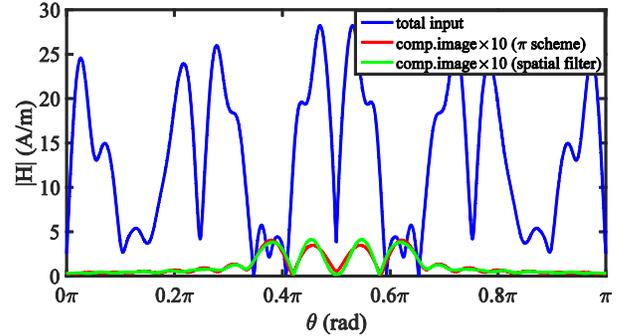

FIG. 6. Comparison of the images obtained by spatial filtering (green) and the Π scheme (red) using a coherent superposition of an auxiliary object and the original object as the total input (blue) for the hyperlens.

Finally, one can anticipate the same or even better resolution by using a thinner hyperlens in conjunction with the loss compensation procedure compared to using a thicker hyperlens alone. At a constant frequency, due to the conservation of momentum, the maximum $k_\theta$ is larger for a thinner hyperlens. This means that evanescent fields appear earlier since $k_r$ needs to be smaller for propagating modes. Therefore, a narrower range of evanescent modes in Fig. 4(a) are converted to propagating modes using a thinner hyperlens. As a result, a thicker hyperlens will have a better resolution. However, by combining the thinner hyperlens with the loss compensation technique presented here, higher resolution achievable through thicker lens alone can be restored at the expense of some magnification. So far, the fabricated hyperlenses in the literature are mainly metal-dielectric layered structures. By making the hyperlens thinner, the number of layers is reduced, consequently reducing the difficulty and cost of fabrication.

### III. CONCLUSIONS

In conclusion, using one fabricated hyperlens as an example, this Letter demonstrates that the plasmon injection scheme for loss compensation in metamaterials can be applied to enhance the hyperlens resolution. To achieve this, the hyperlens transfer function has been numerically calculated. Then in the spatial frequency domain, the raw image spectrum has been multiplied by the inverse of the transfer function. Using this simple post processing, a resolution of $\lambda_0/7$ is achieved, while the image of the objects obtained by the hyperlens alone cannot be resolved. In order to avoid spherical aberration, cylindrical waves and a curved image plane are defined. However, in real application, the source would likely be plane wave and the image plane would be defined by a microscope objective. Additionally, it is more convenient to process intensity information than the complex field requiring both phase and amplitude information, although the latter is still possible [27,28]. In the future work, the above points should be studied to make this loss compensation technique more practical.

**ACKNOWLEDGMENTS**

This work was partially supported by Office of Naval Research (award N00014-15-1-2684) and by the National Science Foundation (NSF) under Grant No. ECCS-1202443. M. S. gratefully acknowledges support from NSF. M. S. gratefully acknowledges support from NSF.

1. J. B. Pendry, "Negative refraction makes a perfect lens," Phys. Rev. Lett. **85**, 3966 (2000).
2. D. R. Smith, D. Schurig, M. Rosenbluth, and S. Schultz, "Limitations on subdiffraction imaging with a negative refractive index slab," Appl. Phys. Lett. **82**, 1506 (2003).
3. X. Zhang and Z. Liu, "Superlenses to overcome the diffraction limit," Nature Mater. **7**, 435 (2008).
4. N. Fang, H. Lee, C. Sun, and X. Zhang, "Sub-diffraction-limited optical imaging with a silver superlens," Science **308**, 534 (2005).
5. Z. Jacob, L. V. Alekseyev, and E. E. Narimanov, "Optical hyperlens: far-field imaging beyond the diffraction limit," Opt. Express **14**, 8247 (2006).
6. B. Wood and J. B. Pendry, "Directed sub-wavelength imaging using a layered metal-dielectric system," Phys. Rev. B **74**, 115116 (2006).
7. Z. Liu, H. Lee, Y. Xiong, C. Sun, and X. Zhang, "Far-field optical hyperlens magnifying sub-diffraction-limited objects," Science **315**, 1686 (2007).
8. J. Rho, Z. Ye, Y. Xiong, X. Yin, Z. Liu, H. Choi, G. Bartal, and X. Zhang, "Spherical hyperlens for two-dimensional sub-diffractional imaging at visible frequencies," Nature Commun. **1**, 143 (2010).
9. J. Sun, M. Shalaev, and N. Litchinitster, "Experimental demonstration of a non-resonant hyperlens in the visible spectral range," Nature Commun. **6**, 7201 (2015).
10. J. Gwamuri, D. O. Guney, and J. M. Pearce, "Advances in plasmonic light trapping in thin-film solar photovoltaic devices," in *Solar Cell Nanotechnology*, A. Tiwari, R. Boukherroub, and M. Sharon, eds. (Wiley, Beverly, 2013), pp. 243–270.
11. K. Aydin, V. E. Ferry, R. M. Briggs, and H. A. Atwater, "Broadband polarization-independent resonant light absorption using plasmonic super absorbers," Nat. Commun. **2**, 517 (2011).
12. V. V. Temnov, "Ultrafast acousto-magneto-plasmonics," Nat. Photonics **6**, 728 (2012).
13. M. I. Aslam and D. O. Guney, "On negative index metamaterial spacers and their unusual optical properties," Progress in Electromagnetics Research B **47**, 203 (2013).
14. M. Sadatgol, M. Rahman, E. Forati, M. Levy, and D. O. Guney, "Enhanced Faraday rotation in hybrid magneto-optical metamaterial structure of bismuth-substituted-iron-garnet embedded-gold-wires," J. Appl. Phys. **119**, 103105 (2016).
15. E. Abbe, "Beiträge zur Theorie des Mikroskops und der mikroskopischen Wahrnehmung," Arch. f. Mikr. Anat. **9**, 413–420 (1873).
16. A. Poddubny, I. Iorsh, P. Belov, and Y. Kivshar, "Hyberbolic metamaterials," Nat. Photonics **7**, 948 (2013).
17. X. Zhang, S. Debnath, and D. O. Guney, "Hyperbolic metamaterial feasible for fabrication with direct laser writing processes," J. Opt. Soc. Am. B **32**, 1013 (2015).
18. D. O. Guney, Th. Koschny, and C. M. Soukoulis, "Reducing ohmic losses in metamterials by geometric tailoring," Phys. Rev. B **80**, 125129 (2009).
19. H. Lee, Z. Liu, Y. Xiong, C. Sun, and X. Zhang, "Development of optical hyperlens for imaging below the diffraction limit," Opt. Express **15**, 15886 (2007).
20. D. O. Guney, Th. Koschny, and C. M. Soukoulis, "Surface plasmon driven electric and magnetic resonators for metamaterials," Phys. Rev. B **83**, 045107 (2011).
21. M. I. Aslam and D. O. Guney, "Surface plasmon driven scalable low-loss negative-index metamaterial in the visible spectrum," Phys. Rev. B **84**, 195465 (2011).
22. M. Sadatgol, S. K. Ozdemir, L. Yang, and D. O. Guney, "Plasmon injection to compensate and control losses in negative index metamaterials," Phys. Rev. Lett. **115**, 35502 (2015).
23. S. Xiao, V. P. Drachev, A. V. Kildishev, X. Ni, U. K. Chettiar, H.-K. Yuan, and V. M. Shalaev, "Loss-free and active optical negative-index metamaterials," Nature **466**, 735 (2010).
24. M. I. Stockman, "Spaser action, loss compensation, and stability in plasmonic systems with gain," Phys. Rev. Lett. **106**, 156802 (2011).
25. W. Adams, M. Sadatgol, X. Zhang, and D. O. Guney, Department of Electrical and Computer Engineering, Michigan Technological University, 1400 Townsend Dr., Houghton, MI 49931-1295, USA, submitted a manuscript called "Bringing the "perfect lens" into focus by near-perfect compensation of losses without gain media."
26. Y. Chen, Y.-C. Hsueh, M. Man, and K. J. Webb, "Enhanced and tunable resolution from an imperfect negative refractive index lens," J. Opt. Soc. Am. B **33**, 445 (2016).
27. T. Taubner, D. Korobkin, Y. Urzhumov, G. Shvets, and R. Hillenbrand, "Near-field microscopy through a SiC superlens," Science **313**, 1595 (2006).
28. J. R. Fienup, "Phase retrieval algorithms: a comparison," Appl. Opt. **21**, 2758 (1982).